\documentclass[aps,pra,superscriptaddress,twocolumn]{revtex4-1}

\usepackage{graphicx}
\usepackage{epstopdf}

\begin{document}

\title{Laser-cooled atoms inside a hollow-core photonic-crystal fiber}

\author{M.~Bajcsy}
\email{bajcsy@stanford.edu}

\author{S.~Hofferberth}
\email{hofferbe@physics.harvard.edu}
\affiliation{Harvard-MIT Center for Ultracold Atoms, Department of Physics, Harvard University, Cambridge, MA 02138}

\author{T.~Peyronel}
\affiliation{MIT-Harvard Center for Ultracold Atoms, Department of Physics, MIT, Cambridge, MA 02139}

\author{V.~Balic}
\affiliation{Harvard-MIT Center for Ultracold Atoms, Department of Physics, Harvard University, Cambridge, MA 02138}

\author{Q.~Liang}
\affiliation{MIT-Harvard Center for Ultracold Atoms, Department of Physics, MIT, Cambridge, MA 02139}

\author{A.~S.~Zibrov}
\affiliation{Harvard-MIT Center for Ultracold Atoms, Department of Physics, Harvard University, Cambridge, MA 02138}

\author{V.~Vuletic}
\affiliation{MIT-Harvard Center for Ultracold Atoms, Department of Physics, MIT, Cambridge, MA 02139}

\author{ M.~D.~Lukin}
\affiliation{Harvard-MIT Center for Ultracold Atoms, Department of Physics, Harvard University, Cambridge, MA 02138}

\date{\today}

\begin{abstract}
We describe the loading of laser-cooled rubidium atoms into a single-mode hollow-core photonic-crystal fiber. Inside the fiber, the atoms are confined by a far-detuned optical trap and probed by a weak resonant beam. We describe different loading methods and compare their trade-offs in terms of implementation complexity and atom-loading efficiency. The most efficient procedure results in loading of  $\sim$30,000 rubidium atoms, which creates a medium with optical depth $\sim$180 inside the fiber. Compared to our earlier study \cite{Lukin2009} this represents a six-fold increase in maximum achieved optical depth in this system.  
\end{abstract}

\pacs{}
\maketitle

\section{Introduction}
Linear and nonlinear light-matter and light-light interactions at few photon levels are essential ingredients for potential applications in areas such quantum communication and quantum information processing \cite{Zeilinger2000}. A key requirement for these is the realization of strong atom-light interaction. Tight confinement of light dramatically increases the atom-light interaction by increasing the electric-field amplitude of single photons. A large variety of systems providing atom-light interaction in confined micro- and nanoscale geometries has been explored in recent years, such as high-Q optical microresonators \cite{Kimble2006,Kimble2008}, tapered optical fibers \cite{Shahriar2008,Hakuta2009,Rauschenbeutel2010}, semiconductor waveguides \cite{Schmidt2007,Lipson2008}, as well as hollow-core photonic-crystal fibers (PCFs) \cite{Russell1999}. 
In particular, geometries based on hollow waveguides, such as those described in  \cite{Schmidt2007, Lipson2008, Russell1999}, offer a platform where the large interaction probability between single photons and single atoms due to tight transverse confinement is unrestricted by diffraction.

Hollow-core PCFs are a special class of optical fibers, where the guided light is confined to an empty central region through a photonic-bandgap effect \cite{Joannopoulos1995}. Atomic or molecular gasses can be inserted into the core of this photonic waveguide, and PCFs filled with various room-temperature atomic or molecular gasses have been used for a wide range of applications, such as spectroscopy \cite{Russell2005b,Henningsen2007,Gaeta2010}, nonlinear optics at low light levels \cite{Russell2002,Gaeta2006,Luiten2007, Benabid2007,Gaeta2009}, and gas sensing \cite{MacPherson2008}.

Here, we present an approach making use of laser-cooled atoms trapped inside such a hollow-core fiber. The confinement prevents atom-wall collisions inside the fiber core \cite{Gaeta2005}, while the Doppler width is smaller than the natural linewidth of atomic transitions. With the flexibility provided by ultracold atomic gasses and quantum optical techniques, atom-atom, atom-photon, and photon-photon interactions in this system have the potential to be engineered in novel ways. For example, we have demonstrated an all-optical switch controlled with less than 1000 photons in this system \cite{Lukin2009}, while recent theoretical proposals \cite{Lukin2005, Lukin2008, Lukin2009b, Lukin2010} predict that controllable nonlinear interaction between single photons can be achieved in the same system for sufficiently large optical depth of the atomic ensemble. In this paper we describe in detail our procedure to load an ensemble of laser-cooled atoms produced by a magneto-optical trap (MOT) into such a hollow-core single-mode photonic-crystal fiber and probe the trapped atoms with resonant light guided by the fiber.  

The use of an optical dipole trap inside a hollow optical waveguide with the goal of guiding trapped cold atoms over macroscopic distances was first theoretically proposed in \cite{Letokhov1993}. Due to the difficulty of guiding light inside low-refractive-index regions, the initial experimental demonstrations done with simple glass capillaries struggled with rapid attenuation of the trap beam \cite{Renn1995}. Additionally, speckle patterns forming as a result of multimode propagation of light inside the capillary would create attractive spots on the capillary walls and cause significant atom loss. In these experiments, the best atom guiding was achieved with evanescent-light fields from blue-detuned laser light injected into the annular glass region of the capillary \cite{Cornell2000}. However, the nature of light propagation in the capillaries made it difficult to attempt efficient nonlinear optical processes in these confined atomic ensembles. The potential of the hollow-core PCF for atom guiding was first demonstrated by Takekoshi and Knize, in whose experiment room temperature atoms were guided through a hollow-core PCF with the help of a red detuned dipole trap \cite{Knize2007}. At around the same time, Christensen \textit{et al.} demonstrated reversible loading of a sodium BEC into a red detuned dipole trap guided by a PCF with a 10 $\mu$m hollow core \cite{Ketterle2008}. Most recently, Vorrath \textit{et al.} report detecting cold rubidium atoms after guiding them through an 88mm-long section of a hollow PCF with a 12 $\mu$m diameter core \cite{Vorrath2010}. 
However, the PCFs used in experiments described in \cite{Ketterle2008} and \cite{Vorrath2010} were not designed to guide light resonant with the atomic transition, and the atoms were not probed by light guided through the fiber. In contrast, the experimental setup introduced here allows for both the optical trapping light and the resonant probing light to be guided by the PCF, enabling us to study interactions between few-photon pulses guided by the hollow-core PCF and atoms trapped inside this fiber. 

\section{Experimental setup}
\begin{figure*}
   \begin{center}
   \begin{tabular}{c}
   \includegraphics[width=14cm]{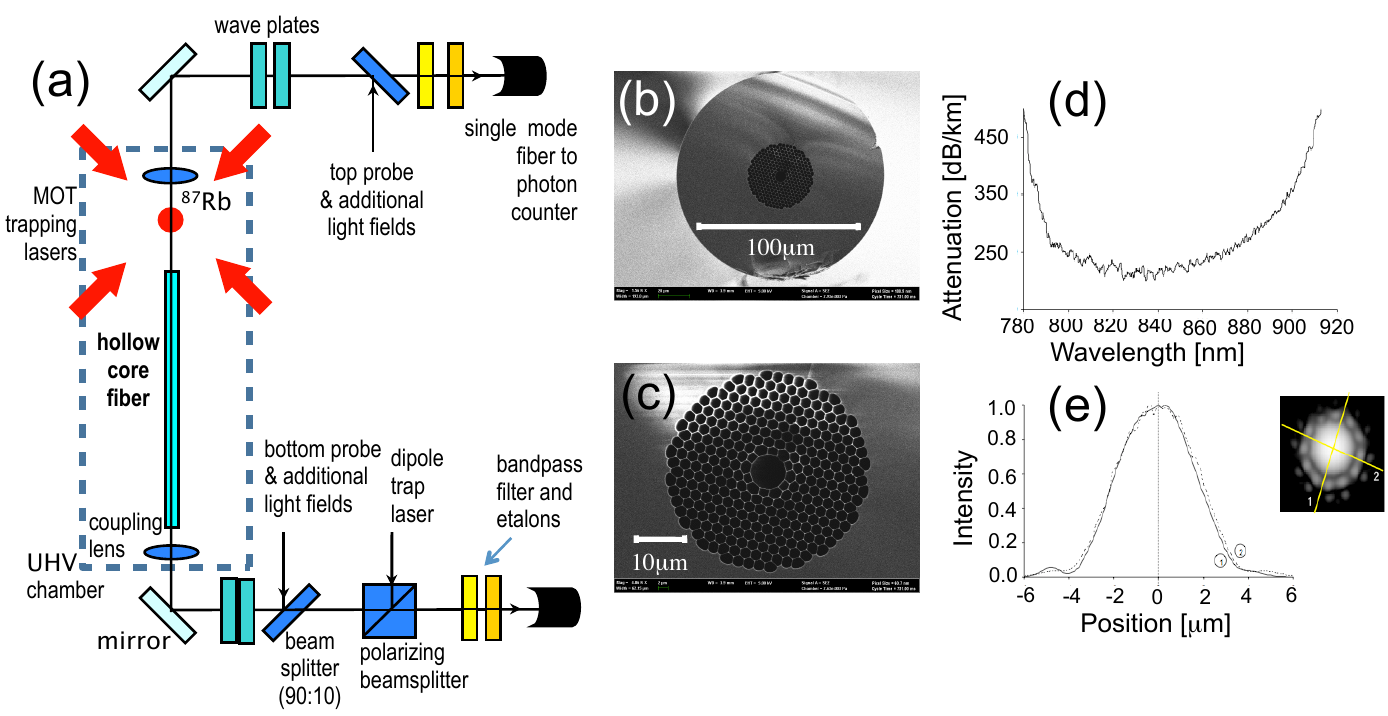}
   \end{tabular}
   \end{center}
   \caption{\label{fig_setup} (a) The schematics of the experimental setup. (b) Scanning electron microscope (SEM) image of a cleaved hollow core photonic crystal fiber, model \textit{HC-800-02}, from Blaze Photonics. (c) Detail of the photonic crystal region with the hollow core in the center. Manufacturer's specifications for: (d) losses of guided mode propagating in the fiber as a function of wavelength and (e) near-field intensity distribution of the guided mode.}
\end{figure*} 

Our apparatus (Fig.~\ref{fig_setup}a) makes use of a $3$\,cm-long piece of single-mode hollow-core PCF vertically mounted inside an ultra-high vacuum chamber. Inside the fiber, the atoms are radially confined by a red-detuned dipole trap formed by a single beam coupled into the fiber from the bottom side. The small diameter of the guided mode allows for strong transverse confinement (trapping frequencies $\omega_t/2 \pi \sim 50-100$\,kHz) and deep trapping potential ($\sim10$\,mK) at guiding-light intensities of a few milliwatts. Since the atoms are attracted toward high light intensity, the diverging beam emerging from the fiber tip creates a potential gradient outside the fiber that attracts cold atoms from the vicinity of the fiber end into the fiber core. During the experiment, a laser-cooled cloud of $^{87}$Rb atoms is collected into a magneto-optical trap, transferred into the vicinity of the upper tip of the PCF, and loaded into the dipole trap guided inside the hollow-core fiber.

\subsection{Fiber mounting structure} 
\begin{figure}
   \begin{center}
   \begin{tabular}{c}
    \includegraphics[width=8.5cm]{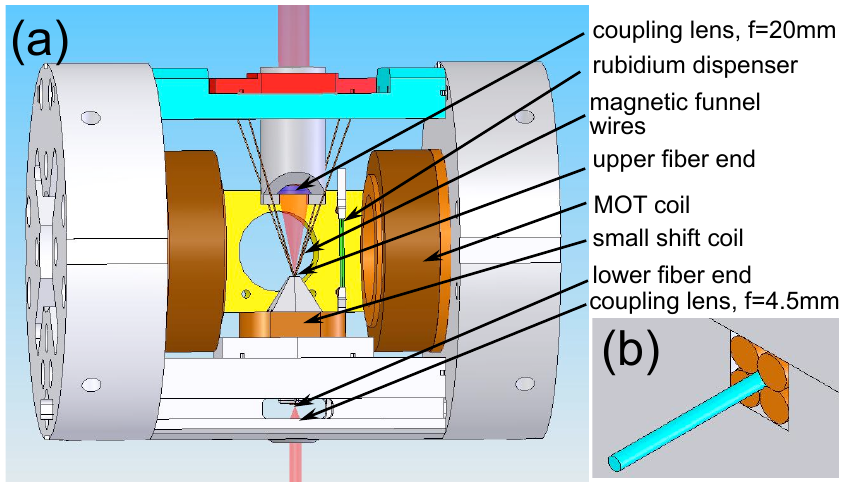}
   \end{tabular}
   \end{center}
   \caption{\label{fig_assembly}(a) Fiber assembly. (b) The anchoring of the fiber inside the mount.}
\end{figure}

The fiber (Fig.~\ref{fig_setup}(b-e)) used in the experiment, \textit{HC-800-02} from Blaze Photonics, has a $7$\,$\mu$m diameter hollow core, and guides light with wavelengths between $780$\,nm and $900$\,nm. The fiber is the centerpiece of a custom-made, ultra-high vacuum compatible assembly mount that includes coupling and imaging optics, as well as magnetic-field generating structures (Fig.~\ref{fig_assembly}(a)).  

The fiber piece is held between four Kapton coated copper wires, which run parallel to the fiber and fan out upwards in an upside-down pyramid configuration above the fiber (shown in Fig.~\ref{fig_assembly}a and \ref{fig_loading}). The diameter of these wires is chosen such that when the wires are packed in a tightly fitting rectangular slit (Fig. \ref{fig_assembly}b), the fiber snugly fits into the space between them. When current of the appropriate polarity is applied to them, the wires act as a magnetic quadrupole guide, focusing the atomic cloud as it is being transferred towards the fiber tip.

In addition to these wires, a parallel pair of coils (main axis horizontal, perpendicular to the fiber) is integrated into the fiber mount. Their symmetry center is located slightly above the fiber tip to create a magnetic quadrupole field for the initial stages of the experiment. 

Besides the current-carrying structures, several optical elements are integrated into the fiber mount as well. Two short-focal-length lenses allow coupling of light into the guided mode of the fiber ($f=20$\,mm for the lens above the fiber, $f=4.5$\,mm for the lens below). These lenses allow us to couple light into the single mode of the PCF with efficiency up to $\sim 40\%$ for light of wavelength 795~nm or longer and up to $\sim 25\%$ for light in the 780-785~nm range. We believe this less than ideal coupling is caused by imperfections in the cleaving of this particular fiber piece, as we observe up to $80\%$ coupling efficiency in other fiber pieces of similar length.
Additionally, a single lens with $f\approx25$\,mm is used for magnified absorption imaging of the atoms near the fiber entrance in the area of the magnetic guide. 

\subsection{Detecting atoms inside the fiber}
To probe the atoms in the fiber, we monitor the transmission of two very-low-power ($\sim 1$\,pW) probe beams with single-photon counters. The beams are coupled into the PCF from either side, as shown in Fig.~\ref{fig_setup}a. After the probes emerge from the fiber, they are collimated by the coupling lenses and then passed through a series of optical filters that separate the probe photons from other light beams coupled into the fiber during the experiments. Finally, the probes are coupled into single-mode fibers connected to photon counters. This last step provides spatial filtering that ensures that only photons propagating in the guided mode of the PCF reach the photon counter.  
The telltale signal of atoms inside the fiber is absorption of the probe light when the probe laser frequency is scanned over an atomic resonance. In the absence of Doppler-broadening the lineshape of such absorption resonances is Lorentzian
\begin{equation}
\label{natural_Lineshape}
T_{nat}=\exp\left(-\frac{\mathrm{OD}} {1+4(\frac{\delta_p}{\Gamma_{e}})^2}\right),
\end{equation}
where $\Gamma_{e}$ is the linewidth of the excited atomic state and $\delta_p = \omega_{p} - \omega_{0}$ is the detuning of the probe laser from resonance. The optical depth $\mathrm{OD}$ is a figure of merit for the strength of the observed absorption. In general, OD depends on the atomic density integrated along the fiber, and the strength of the considered atomic transition. In our experiment, the atoms are confined within the optical trap created by the guided light inside the fiber. Consequently, the radial extent of the atomic cloud is smaller than the beam area of the single-mode probe light beam propagating through the fiber. To get an accurate relation between optical depth and atomic density inside the fiber, we have to take into account the atoms' radial distribution in the probe beam. In particular, an atom at the edge of a beam experiences a smaller electric field and therefore absorbs less light than an atom on the beam's axis. Assuming a Gaussian beam with waist $w_o$ and a radially symmetric atomic density $n(r,z)$, the expression for optical depth on resonance is
\begin{equation}
\label{od3}
\mathrm{OD_{fiber}}={2\over \pi w_o^2}\int_{L_{cloud}}  2\pi\int_0^{r_{core}}n(z,r)c^2_{CG}\sigma_o e^{-{2r^2\over w_o^2}}rdrdz
\end{equation}
where $\sigma_o=\frac{3 \lambda^2}{2\pi}$ is the maximal atomic cross-section, and $c_{CG}$ is the Clebsch-Gordon coefficient for the specific atomic transition being used. In general, (\ref{od3}) reduces to a simple expression that shows that $\mathrm{OD}$ is proportional to the number of atoms $N_{at}$ inside the fiber: 
\begin{equation}
\label{od4}
\mathrm{OD_{fiber}} = \eta \, N_{at} \, {2 c^2_{CG} \sigma_{o}\over \pi w_0^2}.
\end{equation}
The prefactor $\eta$ is given by the radial distribution of atoms in the fiber-confined cloud. The highest value of $\eta$ corresponds to all atoms being localized on the axis of the fiber, in which case $\eta=2$. In the case of a Gaussian radial density distribution $n(r)=n_0 e^{-{r^2\over 2x_0^2}}$, $\eta={2(w_0/2)^2\over x_0^2 +(w_0/2)^2}$. From these considerations it becomes apparent that in our experiment, the optical depth is not solely determined by the number of atoms inside the fiber, but also by the temperature of the atoms, with OD decreasing for higher atomic temperatures. Assuming an atomic temperature $T\sim1$\,mK and using the measured beam waist of guided light inside the fiber $w_0 = 1.9 \pm 0.2 $\,$\mu$m, $N_{at}\sim 100$ atoms inside the fiber create an optically dense medium ($OD= 1$).

\section{Loading procedure} 
\begin{figure*}
   \begin{center}
   \begin{tabular}{c}
    \includegraphics[width=14cm]{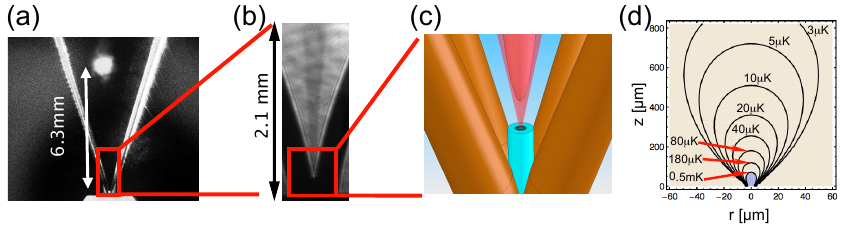}
   \end{tabular}
   \end{center}
   \caption{\label{fig_loading}The loading procedure: (a) Atoms collected in a MOT above the fiber. (b) Absorption image of the atoms in the magnetic-funnel area above the fiber. (c) Once near the fiber tip, the atoms are transferred into a red-detuned dipole trap inside the fiber. (d) Contour plot of the dipole trap potential above the fiber tip resulting from the diverging beam emerging from the fiber tip (located at the origin). The contour labels correspond to a 10~mK deep trap inside the fiber resulting from $\sim$25~mW of 802~nm trap light inside the fiber.}
\end{figure*} 
The starting point of our fiber loading procedure is a standard six-beam magneto-optical trap (MOT) located approximately $6$\,mm above the upper tip of the fiber piece (Fig.~\ref{fig_loading}a). The required light fields are provided by three crossed retro-reflected beams with one-inch diameter, while the magnetic field is realized by the two circular coils inside the vacuum chamber operated in an anti-Helmholtz configuration. During a $\sim1$\,s loading phase we collect about $10^7$ $^{87}$Rb atoms at a temperature of $\sim 100$\,$\mu$K in the MOT from the room-temperature rubidium vapor produced inside the vacuum chamber by a heated dispenser. Following this step, the magnetic fields are ramped up over a period of $40~$ms to compress the cloud, the frequency of the trapping beams is moved from the initial 15 MHz off-resonance to 50-60 MHz detuning, and their power is reduced by a factor of $\sim 4$. Finally, the magnetic fields are shut off, and the atomic cloud is allowed to slowly expand for $10~$ms in the optical molasses of the intersecting beams as it undergoes polarization gradient cooling. This last cooling step lowers the cloud temperature to $\sim 40~\mu$K.

After the laser-cooling stage, we transfer the atoms downwards into the vicinity of the fiber tip (Fig.~\ref{fig_loading}b, c) from where they are loaded into the fiber. Over the course of the experiment, we have implemented different procedures for this transfer, which are described in the following.

\subsection{Magnetic funnel guiding}
The original transfer procedure is based on magnetic guiding of the atoms. After the initial cooling stages in the MOT and optical molasses are completed, the atoms are optically pumped into the $\left|F=2,m_F=2\right>$ state and then transferred into a magnetic quadrupole trap formed by the same coils which provide the MOT field. This trap is then adiabatically shifted towards the fiber tip by adding a vertically-oriented homogeneous offset field, which displaces the zero-field center of the quadrupole trap. In addition, current in the magnetic funnel wires is turned on, creating a transverse quadrupole field, in which the gradient increases with decreasing distance from the upper fiber tip. At the fiber tip this transverse gradient reaches $\sim 6$\,kG/cm, resulting in strong radial compression of the magnetic trap (Fig.~\ref{fig_loading}b). The complete transfer of the magnetic trap towards the fiber takes place over the course of $45$\,ms. This brings the atoms within a few hundred micrometers of the fiber tip. During the transfer stage, the fiber-guided dipole trap is turned on, so that when the atoms start approaching the fiber face, they are captured by the expanding beam of the dipole trap and pulled into the hollow core of the PCF(Fig.~\ref{fig_loading}c, d). At the end of the transfer, all magnetic fields are shut off and the atoms are probed. With this method we observe the loading of up to $\sim 10^4$ atoms into the fiber, equivalent to a maximum $\mathrm{OD}\sim 50$.

While this procedure loads atoms into the fiber reliably, it turns out to have significant drawbacks. When current is pulsed through the funnel wires, the resulting heat pulse causes the fiber tip to shake slightly. Additionally, the cumulative heat of the repeated experimental cycles causes drifts in the overall fiber coupling efficiency. This requires the system to run for about two hours before the fiber position stabilizes and after that the funnel needs to be cycled constantly to maintain the steady-state temperature of the fiber mount. 

\subsection{Hollow-beam atomic guide}
\begin{figure}
   \begin{center}
   \begin{tabular}{c}
   \includegraphics[width=8.5cm]{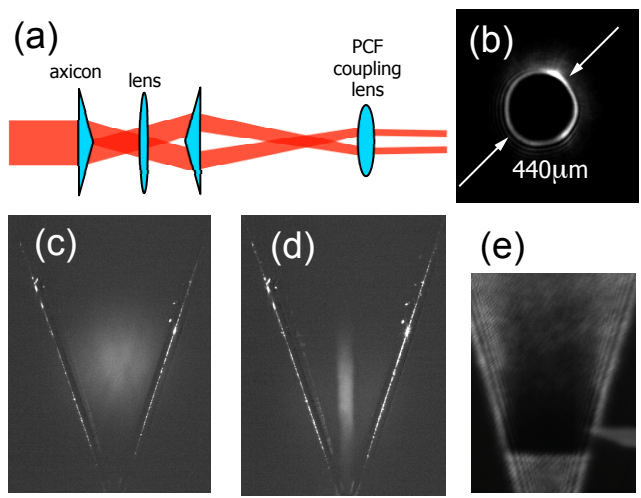}
   \end{tabular}
   \end{center}
   \caption{\label{fig_hollowguide}Hollow-beam atomic waveguide. (a) Schematics of the optics used for the hollow-beam generation. (b) CCD image of the hollow-beam intensity distribution about 1~mm above the fiber face. (c) Fluorescence image of the freely expanding atomic cloud 20~ms after its release from the optical molasses. (d) Fluorescence image of the atomic cloud guided by the blue detuned hollow beam 20~ms after the optical molasses beams are turned off. (e) Absorption image of the atoms collected in the hollow beam guide $\sim 1~$mm above the fiber tip. Here, the hollow beam is intersected by a blue detuned Gaussian beam focused by a cylindrical lens into a sheet.} 
\end{figure} 

The problems associated with the pulsed currents required for the magnetic transfer led us to the development of an all-optical transfer method. Instead of capturing the initial MOT in a magnetic trap we now confine it transversally by an optical guiding potential. This atomic guide is based on a hollow-beam blue-detuned dipole trap. The hollow beam is generated using a combination of lenses and axicons (conical lenses) sketched out in Fig.~\ref{fig_hollowguide}a and described in more detail in \cite{Bajcsy2009}. This setup allows us to generate a vertical hollow beam that is close to collimated both in diameter and wall thickness in the region between the MOT site and $\sim1~$mm above the fiber tip (Fig~\ref{fig_hollowguide}b). The idea behind this particular lens combination is to turn ``inside out'' an axicon-generated quasi-Bessel beam, which leads to an excellent suppression of light in the hollow part of the resulting beam \cite{Song1999}. One practical constraint in our implementation of this optical guide is that it has to pass through the $f=20\,$mm collimation lense above the fiber. Consequently, the other optical elements, located outside the vacuum chamber, have to be matched to this lens. This leads to a  combination of optics consisting of two 175$^{\circ}$ axicons (Greyhawk Optics) and a 75~cm focal-length lens between them. The hollow beam shape is fine-tuned by adjusting the collimation of the input Gaussian beam.

The atom-guiding performance of the blue-detuned hollow beam generated with this setup can be seen in Fig.~\ref{fig_hollowguide} c, d and e. The hollow beam with $P\approx 40~$mW and $\lambda\approx 780.20~$nm is turned on at the end of the atom cooling stage, and the atoms are then allowed to free-fall towards the fiber. Comparison of the fluorescence images of the freely expanding MOT (Fig.~\ref{fig_hollowguide} c) and the optically confined cloud (Fig.~\ref{fig_hollowguide} d) shows how the hollow guide increases the atomic density in the area above the fiber tip, by preventing atoms from escaping from this region. On the other hand, it can be seen that the cloud still expands freely in the vertical direction. To also decrease the size of the atomic cloud in this direction, we add a blue-detuned light sheet perpendicular to the fiber $\sim 1\,$mm above the fiber tip. This closes off the optical trap in the vertical direction, creating a cup-like potential together with the hollow guide, in which the atoms are collected close to the upper fiber tip (Fig.~\ref{fig_hollowguide}e). Once the atoms have accumulated in this cup, the light sheet is turned off and the atoms again fall freely towards the fiber where they are captured by the in-fiber dipole trap. With this method we load $\sim 3\times10^4$ atoms into the fiber, which results in a maximum  $\mathrm{OD}\sim 180$.

\subsection{Free fall}
It is interesting to note that we can also load atoms into the fiber by simply releasing the MOT and letting the atoms fall completely unrestricted. In this case, we observe up to $\sim 5000$ atoms in the fiber, which is within the same order of magnitude as the results achieved by the other transfer methods. This is due to the fact that all our transfer methods are purely adiabatic, i.e. there is no additional cooling of the atoms after the optical molasses stage. Consequently, any transverse compression of the atomic cloud will result in an increase of its temperature, which in turn reduces the chance of individual atoms being loaded into the fiber dipole trap. 
In particular, most of the transverse compression happens in the last $<100\,\mu$m above the fiber, where the potential from the fiber guided dipole trap becomes significant (Fig.~\ref{fig_loading}d). Only a fraction of the atoms passes through this area into the fiber. The majority has too much kinetic energy which results in trajectories that do not end inside the fiber, but instead lead the atoms back upwards, anagolous to the angular momentum barrier in a magnetic bottle trap. To increase the number of atoms passing this barrier, optical cooling during this final compression stage is required. In our current setup, the magnetic funnel wires block the optical access to this region, which prevents us from implementing this improvement.

\section{Atoms inside the fiber} 
After the end of the transfer stage, we wait $\sim$5\,ms before we perform the actual in-fiber experiments, which allows the captured atoms to move into the fiber, and atoms remaining above the fiber to expand sufficiently to not cause any residual absorption. Additionally, if magnetic guiding is used in the transfer process, this step gives the transient magnetic fields time to decay. Typically, the duration of the actual experiments ranges from $100$ to $400$\,$\mu$s. The whole cooling, trapping, and data collection cycle is repeated every $1.5$\,s. For the probe transmission scans such as those shown in Figs.~\ref{fig_transmission}a and \ref{fig_OD}, each data point corresponds to a single run of the experiment (or multiple runs in the case of data averaging). Between these runs, the frequency of the probe laser is changed, and a new atomic sample is prepared in the fiber. 

For the dipole trap inside the fiber we couple $P=25\,$mW of light with wavelength $\lambda=802$\,nm into the fiber. This provides a trap depth of $\sim 10$\,mK, while the detuning from the rubidium lines is sufficient to allow for easy optical filtering and negligible residual heating during the experiments.

\subsection{Modulated dipole trap}
	\label{subsec:modulation}
   \begin{figure}
   \begin{center}
   \begin{tabular}{c}
   \includegraphics[width=8.5cm]{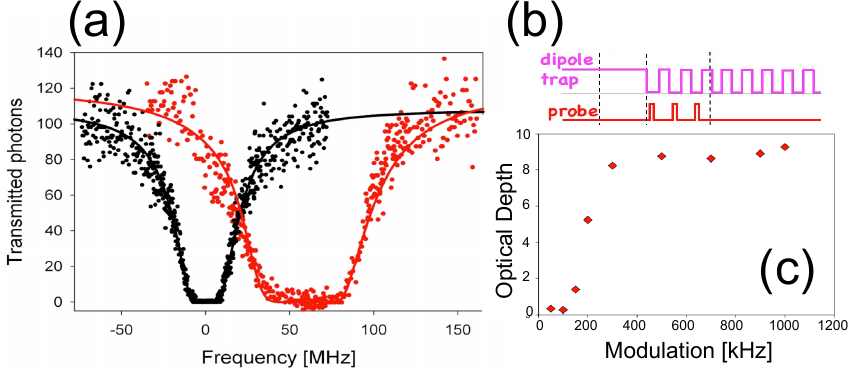}
   \end{tabular}
   \end{center}
   \caption{\label{fig_transmission}Atoms inside the fiber: (a) Transmission of the fiber as a function of probe frequency with constant dipole trap (broad red data curve centered at $\sim60\,$MHz) and with modulated dipole trap (narrower black data curve centered at $0\,$MHz). (b) Modulation scheme for probe and dipole trap: The probe is broken into $\sim 100$ short pulses that interact with the atoms when the dipole trap is off. (c) Detected optical depth of the atomic cloud inside the fiber as a function of the dipole trap modulation frequency.}
\end{figure} 
   
Absorption profiles associated with atomic resonance lines are the signature of the interaction between atoms and light guided through the PCF. If the atoms are probed while they are confined in the in-fiber dipole trap, we observe a unique absorption profile as shown in Fig.~\ref{fig_transmission}a. The dipole trap introduces a power-dependent, radially varying AC-Stark shift \cite{Grimm2000}, which results in broadening and a frequency shift of the absorption profile (red data points in Fig.~\ref{fig_transmission}a). 
For actual experiments, we usually want to avoid this broadening of the atomic transition. To achieve this, we apply a synchronous square-wave modulation of the dipole trap and the probe beam with opposite phase (Fig.~\ref{fig_transmission}b) at a rate much higher than the trapping frequency. This results in a time-averaged trapping potential that is still sufficiently deep to confine the atoms inside the fiber. On the other hand, during the off-times of the dipole trap, we can perform optical experiments with the atoms exhibiting their field-free atomic structure. When using this technique and scanning the probe laser over a particular hyperfine transition, we observe a narrowed absorption profile as shown in figure \ref{fig_transmission}a (black data points). The shape of this resonance is completely determined by the natural line profile of the transition (equation (\ref{natural_Lineshape})). There is solely homogeneous broadening due to the large optical depth. The exact frequency of the trap light modulation is adapted based on the experiments performed. Slower modulation increases the off-times of the trap light, providing more time for individual experiments. Comparing the observed $\mathrm{OD}$ for different modulation frequencies (Figure~\ref{fig_transmission}c) shows that the minimum modulation we can use before we observe loss of atoms is $\sim 300\,$kHz.

\subsection{Number of atoms inside the fiber}
\begin{figure}
   \begin{center}
   \begin{tabular}{c}
   \includegraphics[width=\columnwidth]{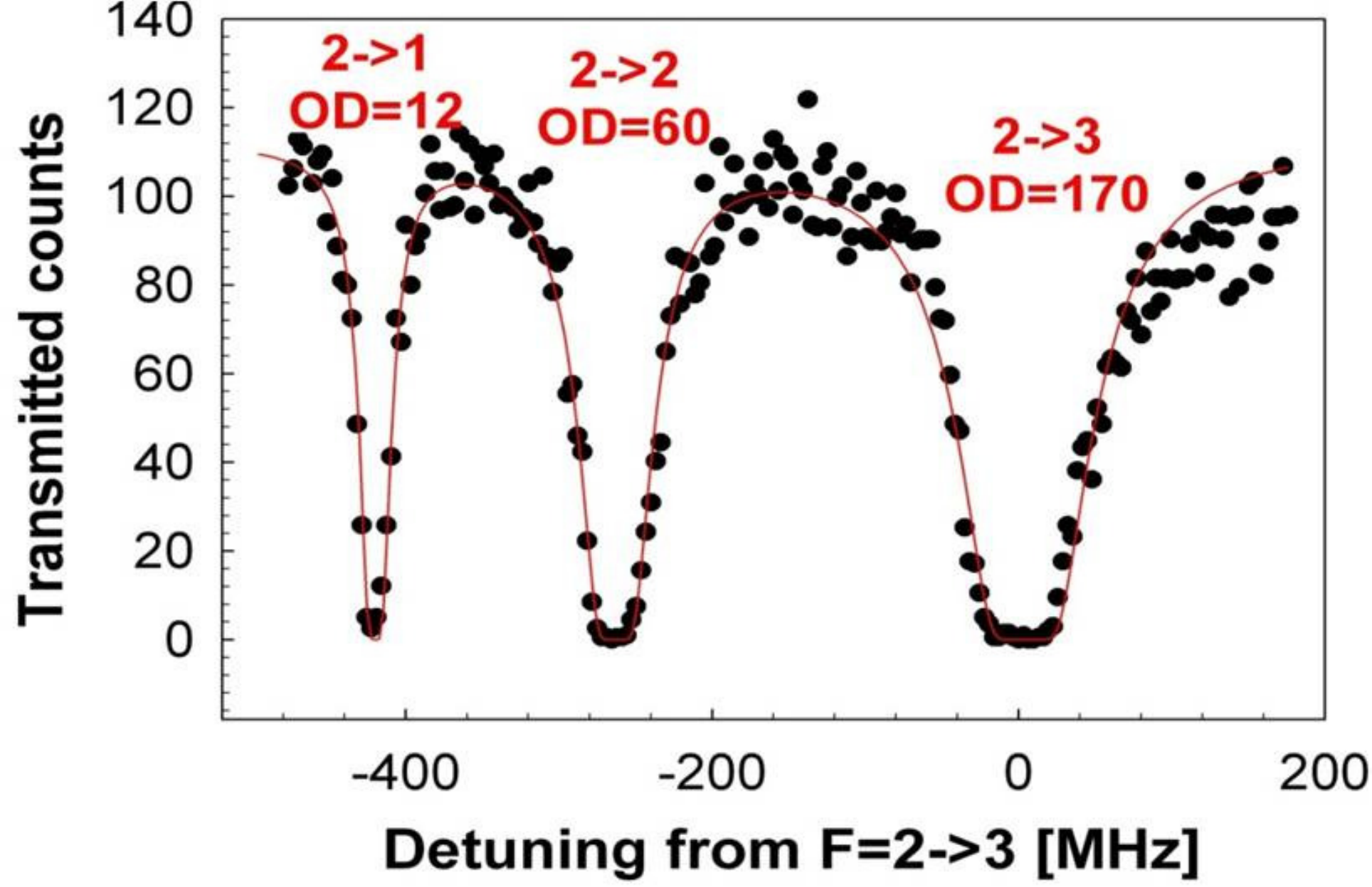}
   \end{tabular}
   \end{center}
   \caption{\label{fig_OD} Frequency scan over the $D_2$ line of $^{87}$Rb for $\sim 26000$ atoms loaded inside the fiber. The atoms are optically pumped into the $F=2$ state and then probed with linearly polarized light over the three transitions accessible from this state, $F=2 \to F'=1$ (left), $F=2 \to F'=2$ (middle), and $F=2 \to F'=3$ (right). The different observed optical depths agree well with a number of loaded atoms of $N_{at}\sim 30000$ when the relative strengths of these transitions are taken into account.} 
\end{figure} 

\begin{figure}
   \begin{center}
   \begin{tabular}{c}
   \includegraphics[width=8cm]{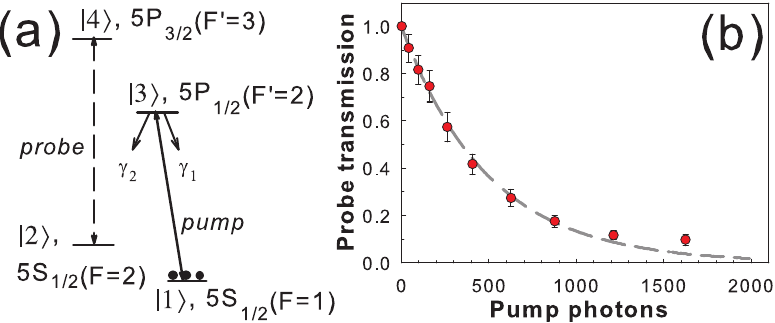}
   \end{tabular}
   \end{center}
   \caption{\label{fig_saturation} (a) Calibration of the number of atoms inside the fiber by counting the number of photons required to transfer the amount of atoms required to cause $\mathrm{OD}=1$ on a given probe transition. (b) With $\gamma_1\approx \gamma_2$, fitting an exponential (dashed line) to the measured data (red dots with error bars) yields $\eta\approx0.42$ as the value of the prefactor from equation (\ref{od4}).} 
\end{figure}

Figure~\ref{fig_OD} shows a typical absorption scan when the atoms are initially prepared in the $F=2$ hyperfine ground state and the probe laser is scanned over the $D_{}2$ transition while the dipole trap is turned off. The maximum optical depth $\mathrm{OD}\sim180$ is observed on the $F=2 \to F'=3$ cycling transition when we use the all-optical loading procedure. While this number grows monotonically with the number of atoms loaded, we cannot straightforwardly extract how many atoms are inside the fiber from the measured OD. To measure this quantity $N_{at}$ we perform a nonlinear saturation measurement based on incoherent population transfer in our mesoscopic atomic ensemble. The transmission of a probe beam coupled to a cycling atomic transition, $\vert 2\rangle \rightarrow \vert 4\rangle$, is controlled via an additional pump beam transferring atoms from an auxiliary state $\vert 1\rangle$ into $\vert 2\rangle$ (Fig. \ref{fig_saturation}a). Initially, the state $\vert 2\rangle$ is not populated and the system is transparent for the probe beam. The incident pump beam, resonant with the $\vert 1\rangle \rightarrow \vert 3\rangle$ transition, is fully absorbed by the optically dense atom cloud, thereby transferring atoms into the $\vert 2\rangle$ state, where they then affect the propagation of the probe beam. As demonstrated in Fig. \ref{fig_saturation}b, we achieve a $50 \%$ reduction of the probe transmission with only $300$ pump photons. This corresponds to $\sim 150$ atoms being transferred into the $\vert 2\rangle$ state, which yields the desired proportionality constant between optical depth on the considered transition and number of atoms inside the fiber. From this number we can calculate this conversion factor for any atomic transition by comparing its strength to the reference cycling transition.

\subsection{Experimental verification of atom loading}
\begin{figure}
   \begin{center}
   \begin{tabular}{c}
     \includegraphics[width=\columnwidth]{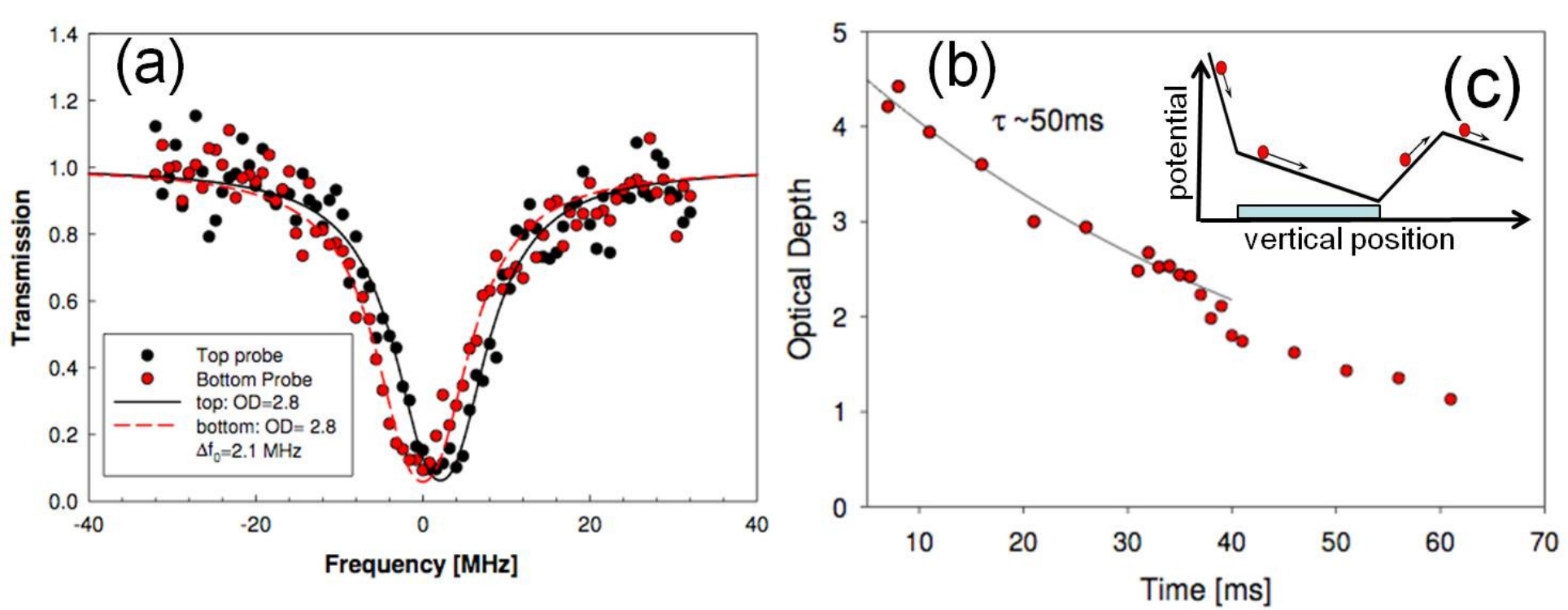}
   \end{tabular}
   \end{center}
   \caption{\label{fig_dopplershift} Atoms inside the fiber: (a) Doppler shift of the falling atom cloud observed by using two probes propagating through the fiber in opposite directions. (b) Lifetime of the atoms inside the fiber. Inset: A qualitative sketch of the potential along the fiber axis experienced by the atoms as they load into the fiber-guided dipole trap.}
\end{figure}

From the absorption scans we cannot infer with absolute certainty that the atoms are inside the fiber. Atoms trapped in some local potential minimum just outside the fiber would create an almost identical absorption signal. Such a minimum could form due to stray trap light at the fiber tip if the fiber is not cleaved properly, for example. Due to the magnetic funnel wires surrounding the fiber, we cannot obtain a direct image of the atoms inside or right above the fiber, which would let us determine their position exactly. 
For the number of atoms we load into the fiber, detection of atoms exiting at the lower end is challenging, as we would need to detect very few atoms at relatively high temperature ($\sim 1\,$mK) in free space. While this can be accomplished, e.~g., by using an image intensified CCD camera \cite{Vorrath2010} or a channel electron multiplier \cite{Knize2007}, we instead confirm that the atoms are loading into the fiber core by measuring their vertical velocity over an extended period of time. For this, we deploy two probes propagating in opposite directions through the fiber simultaneously. The observed absorption profiles show a distinct difference between the frequency centers of the two profiles (Fig.~\ref{fig_dopplershift}a) after the atoms are loaded into the dipole trap. This frequency difference is the result of the Doppler shift caused by the vertical motion of the atomic cloud. In the case shown in figure~\ref{fig_dopplershift}a, the observed shift corresponds to the cloud moving downwards with velocity $0.82$\,$\mathrm{m/s}$. This shift can be observed for times exceeding $30~$ms, which means the atoms have to be moving downward inside the fiber. 

Additionally, we can control the velocity of the atoms inside the fiber in two ways. In the first way, we can simply change the depth of the fiber-guided dipole trap, which modifies the kinetic energy picked up by the atoms when they are loaded into this trap. The second way is based on coupling into the fiber an additional weak ($\sim 1\,$nW) nearly-resonant  red-detuned upward or downward propagating 'push' beam, which allows us to alter the overall velocity as well as the direction of movement of the atomic cloud inside the fiber. The light pressure from this beam -- depending on the beam's direction, intensity, and duration -- can speed up, slow down, stop, or reverse the direction of the movement of the atoms inside the fiber.        

\subsection{Atom-wall interactions} 
Due to the small size of the hollow core, the possibility that attractive forces from the core wall cancel the dipole trap potential has to be considered. However, comparing the attractive Casimir-Polder potential $C_4 /r_{wall}^4$ created by the fiber wall, with the nominal value of  $C_4= 8.2\times 10^{-56} J m^4$ \cite{Babb1997, Vuletic2004}, with the dipole trap potential shows that attractive forces become dominant only for distances less than $\sim 100\,$nm and that the reduction of total depth of the optical trap is negligible. Therefore, for a fiber with a  $7~\mu$m diameter core, the shape or depth of a red-detuned dipole trap potential in the central area is not noticeably affected by the nearby walls.

\subsection{Lifetime of atoms inside the fiber}
Once inside the fiber, the atoms are confined by  the red detuned dipole trap only in the radial direction, while in the vertical direction they experience a free fall until they reach the lower end of the fiber. An example of the measured optical depth of the falling atomic cloud as a function of time is plotted in figure \ref{fig_dopplershift}b. Here, the zero on the time axis corresponds to the instant when the optical depth in the fiber is the largest. In this measurement, each point on the graph corresponds to a newly loaded atomic cloud for which the dipole trap was kept on continuously until the point in time when the OD was measured using the modulation described above. The `kink' in the data near 40~ms corresponds to the free-falling atoms reaching the lower end of the fiber. Up to this point the atoms decay out of the dipole trap exponentially with a time constant of $\sim 40~$ms. We explain the data after the `kink' as part of the atomic cloud leaving the fiber and part of the cloud reflecting back from the potential change associated with the dipole-trap beam being coupled into the lower end of the fiber. 

As the atoms move inside the fiber, they are lost from the dipole trap mostly through two mechanisms. The first and more obvious one is caused by collisions with the background gas present due to imperfect vacuum within the PCF core. While the vacuum pressure is $10^{-9}$ torr or less within the general volume of our vacuum chamber, the small diameter of the fiber core and the associated pumping speeds should result in a significantly higher background gas pressure inside the fiber. Considering the expansion of a room temperature gas in a one-dimensional tube with diameter $7\,\mu$m, we can estimate that the pressure inside the fiber reaches $10^{-6}$ torr after about one day of pumping, while getting down to $10^{-8}$ torr takes a little more than a month. This model neglects possible outgasing from the fiber walls, which will increase the obtainable steady-state pressure. From our observations, background gas collisions inside the fiber are not the main limitation on our observed lifetimes. Instead, we identify a second mechanism leading to faster atom loss, which originates from the presence of higher-order guided modes propagating through the PCF. These modes are present in all single-mode fibers, as their excitation during beam coupling from free space into fiber is virtually inevitable. However, these modes propagate with losses significantly larger than those of the fundamental mode and generally die away when propagating through fiber pieces longer than $\sim1$\,m. For a short piece of fiber, like the one used in the experiment, the losses of the higher order modes will not be sufficient to suppress them.  These higher-order spatial modes interfere with the fundamental mode and create a transverse as well as longitudinal variation of the dipole trap potential. Since this is an interference effect, even small amounts of power propagating in higher modes can lead to a significant modulation of the original potential. These potential variations couple the longitudinal velocity of the atoms, which is quite large as the atoms gain kinetic energy from falling into the dipole trap potential (Fig.~\ref{fig_dopplershift}b, inset), to the radial motion of the atoms. This coupling heats the atoms in the transverse direction and ejects them from the dipole trap. As a result, we have observed a decay constant of the atoms inside the fiber ranging from 100~ms down to 10~ms depending on position and the velocity of the atomic cloud inside the fiber. In particular, we observe the longest lifetimes when the atoms are completely stopped inside the fiber, while we register a reduction of lifetime when the atoms move both up or down inside the fiber. 

\subsection{Length of the atomic cloud inside the fiber}
In general, during the loading procedures the atoms arrive at the upper tip of the fiber with velocities of $\sim 0.35~$m/s over time intervals $\sim 10~$ms. Once they fall into the dipole trap potential, their velocity will reach up to $\sim 1.4~$m/s, depending on the depth of the dipole trap inside the fiber. Consequently, during the $\sim 10~$ms when the atoms are entering the fiber, the front edge of the atomic cloud is moving at $\sim 1.4~$m/s, while its rear edge is moving at $\sim 0.35~$m/s. Based on this, we can estimate the minimum length of the atomic cloud inside the fiber to be $\sim 1~$cm.

\subsection{Temperature of the fiber-confined atomic cloud}
\begin{figure}
   \begin{center}
   \begin{tabular}{c}
   \includegraphics[width=6cm]{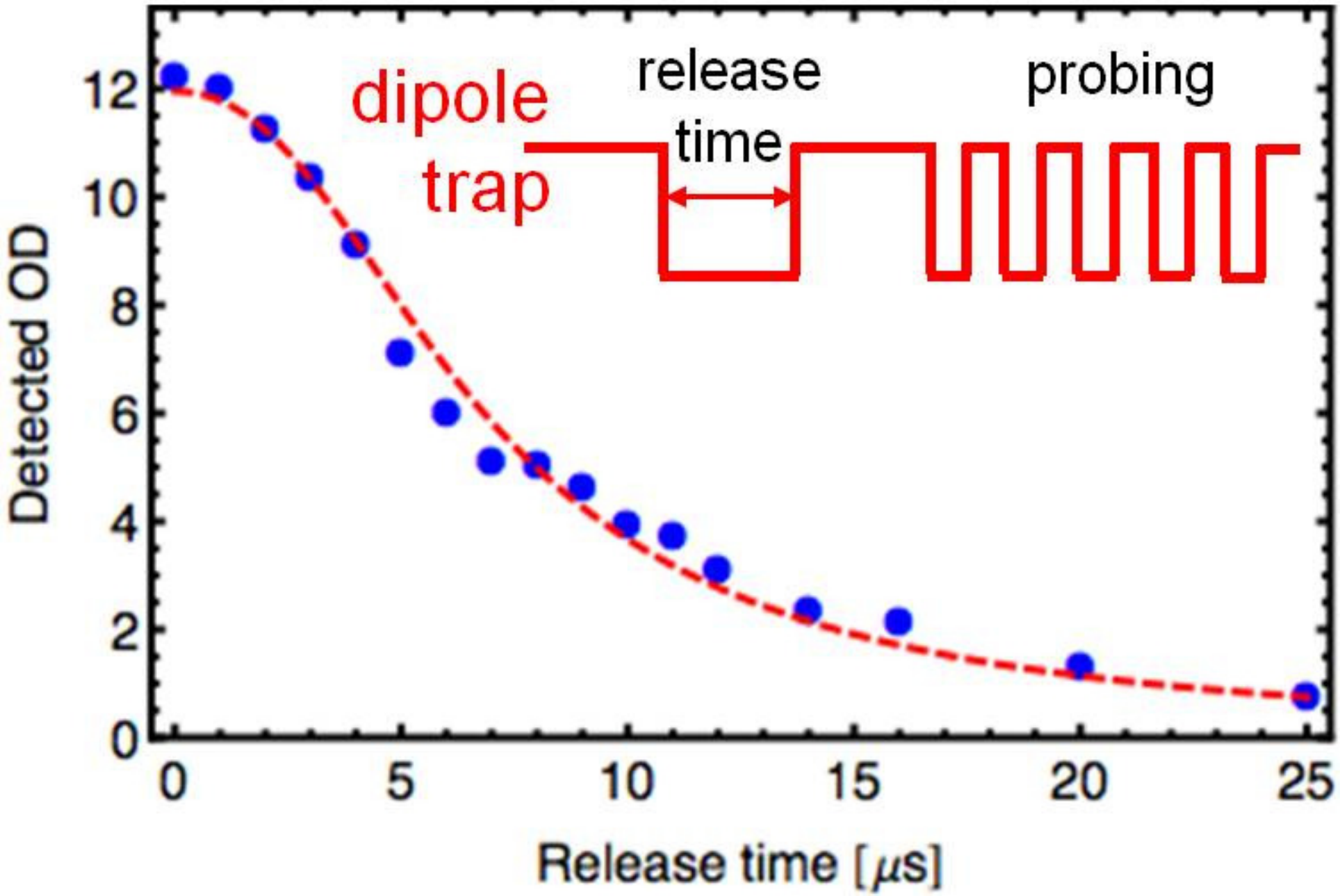}
   \end{tabular}
   \end{center}
   \caption{\label{fig_temperature}Temperature estimate from a TOF measurement: (a) The time sequence diagram: After the atoms are released from the trap and then recaptured, the trap is modulated for the OD measurement. (b) Optical depth of the recaptured cloud as a function of the release time.  The dashed red line represents a fit based on the Gaussian cloud expansion model.}
\end{figure}   

The tight confinement of the fiber-guided dipole trap will increase the temperature of the atoms compared to the initial $40\,\mu$K outside the fiber. To estimate the temperature of the atoms inside the fiber, we perform a time of flight (TOF) measurement. We shut off the dipole trap and let the cloud expand, then turn the trap on again to recapture atoms that have not collided with the wall (Fig. \ref{fig_temperature}a). After this, we modulate the dipole trap in the usual manner and measure the optical depth of the recaptured atoms. 
Assuming a Gaussian distribution of the atoms in the radial direction $n(r)\sim e^{-(r/r_o)^2}$, the optical depth of the recaptured cloud as a function of the release time $\tau_{r}$ is given by
\begin{equation}
\label{recapture}
\mathrm{OD}(\tau_{r}) \approx \mathrm{OD}_0 \left(1-\exp \left[\frac{-\left({R_{core}\over r_o}\right)^2}{1+\left({v_o\over r_o}\right)^2\tau_{r}^2}\right]\right)
\end{equation}
where $v_o=\sqrt{{2kT_{r}\over m_{Rb}}}$ and $R_{core}$ is the radius of the PCF core. We extract the temperature of the cloud by fitting (\ref{recapture}) to a set of release and recapture data with $A=\left( {R_{core}\over r_o}\right)^2$ and $B=\left( {v_o\over r_o}\right)^2$ as the fit parameters, as shown in Fig.~\ref{fig_temperature}b. The fit to this data set yields $T_{r}\approx1.6~$mK and $r_o\approx2.2~\mu$m. The latter corresponds to the prefactor from equation (\ref{od4}) being $\eta\approx0.31$, which roughly agrees with the value obtained independently from the measurement of the number of atoms inside the fiber (Fig. \ref{fig_saturation}). 

\section{Discussion}
Using the hollow-beam atomic guide, we achieve an optical depth of 180 inside the hollow core of the PCF. This number is limited by the peak phase space density $\rho_{MOT}= n_0 \lambda_{dB}^3 \approx10^{-7}$ of the initial MOT, where $n_0$ is the peak atomic density inside the MOT, and $\lambda_{dB}$ is the de Broglie wavelength of the atoms. Inside the fiber, the peak phase space density is given by $\rho_{fiber} = N_{at} (\frac{\hbar \omega_\perp}{k_B T})^2 \frac{\lambda_{dB}}{L}$, where $\omega_\perp$ is the transverse trap frequency, and $L$ is the longitudinal extent of the cloud inside the fiber. Using the measured temperature $T\sim1$\,mK, cloud length $L=1$\,cm, $\omega_\perp=2\pi 50$\,kHz, we obtain $\rho_{fiber} =10^{-7}$, which is identical to the original peak phase space density in the MOT, showing that our transfer is essentially adiabatic. The essence of our current loading implementation is that we rely on capturing the lower tail of the Maxwell-Boltzmann velocity distribution of atoms from the MOT by the fiber-coupled dipole trap. Currently, we only ensure that atoms in this tail are brought into the capture radius of the fiber trap. Some increase in the number of loaded atoms can be obtained by increasing the MOT size, but we observe this to become marginal for MOTs containing more than $\sim 2\times10^7$ atoms.  To substantially increase the loading efficiency, and thereby the number of loaded atoms, one has to increase the phase space density of the atomic cloud by additional cooling of the atoms before or during the transfer into the fiber. In the most extreme case this means loading a Bose-Einstein condensate into the fiber. This was successfully demonstrated in \cite{Ketterle2008}, but leads to a severe decrease in the repetition rate of the experiment in addition to an increased technical complexity of the apparatus. Optical cooling of the atomic cloud in the vicinity of the fiber tip once the atomic cloud has been significantly compressed is prevented by magnetic-funnel wires blocking the optical access. Since our current loading scheme does not rely on these wires anymore, they can be removed, allowing the fiber tip to be optically accessible.
 
\section{Conclusion} 
In summary, we have described a procedure for loading laser-cooled atoms into a single-mode hollow-core photonic waveguide and trapping them sufficiently long to perform optical experiments. We have thoroughly characterized the properties of the atoms inside the fiber, such as lifetime, temperature, and basic interactions with resonant light. With the fiber acting as an atomic guide, the tight confinement of both photons and atoms inside the fiber results in an increased probability of single photon-single atom interaction and, additionally,  the photons can interact with an optically dense atomic ensemble without being limited by diffraction. This makes the system an excellent candidate for nonlinear optics at very low light levels \cite{Lukin2009},  while the large optical depth achieved in our system makes it ideally suited for the implementation of schemes for enhancing optical nonlinearities and creating effective photon-photon interactions \cite{Lukin2005, Fleischhauer2008b, Lukin2008, Fleischhauer2009b, Fleischhauer2010, Gorshkov2010}.  Additionally, we expect this system to be beneficial for applications and experimental studies in a number of other areas, such as physics of one-dimensional systems \cite{Sorensen2010}, quantum nondemolition measurements \cite{Rosenblum2010}, Dirac dynamics \cite{Unanyan2010},  quantum simulation \cite{Hartmann2010}, and quantum telecommunication \cite{Brask2010}.

%

\end{document}